\documentclass[aps,prd,preprintnumbers,superscriptaddress,nofootinbib,floatfix,twocolumn,notitlepage]{revtex4-2}
\usepackage{graphicx}  % needed for figures
\usepackage{dcolumn}   % needed for some tables
\usepackage{bm}        % for math
\usepackage{epsfig,amsmath,amssymb,verbatim,mathrsfs,array,layout,textcomp,amssymb,latexsym,slashed}
\usepackage{xcolor}
\usepackage{siunitx}
\usepackage[colorlinks=true,citecolor=blue,urlcolor=blue,linktocpage=true,
linkcolor=blue]{hyperref}
\usepackage[utf8]{inputenc}
\usepackage{multirow}
\usepackage{amsthm}
\usepackage{thmtools}
\usepackage[capitalize,noabbrev]{cleveref}

\usepackage[normalem]{ulem}

\newcommand{\eV}{\mathrm{eV}}
\newcommand{\meV}{\mathrm{meV}}

\newcommand{\MeV}{\mathrm{MeV}}
\newcommand{\GeV}{\mathrm{GeV}}

\newcommand{\MP}{M_\mathrm{Pl}}

\def\beq{\begin{equation}}
\def\eeq{\end{equation}}
\def\beqa{\begin{eqnarray}}
\def\eeqa{\end{eqnarray}}

\begin{document}

%%%%%%%%%%%%%%%%%%%%%%%%%%%%%%%
\title{Natural Phantom Crossing from Axion-WIMP Interactions}
 
\preprint{LAPTH-046/26}
%%%%%%%%%%%%%%%%%%%%%%%%%%%%%%%

%%%%%%%%%%%%%%%%%%%%%%%%%%%%%%%
%
\author{C\'edric Delaunay}
\email{cedric.delaunay@lapth.cnrs.fr}
\affiliation{Laboratoire d'Annecy de Physique Th\'eorique, CNRS--USMB, 9 chemin de Bellevue, 74940 Annecy, France}

\author{Seung J. Lee}
\email{sjjlee@kias.re.kr}
\affiliation{School of Physics, KIAS, 85 Hoegi-ro, Dongdaemun-gu, Seoul 02455, Korea}
\affiliation{Quantum Universe Center, KIAS, 85 Hoegi-ro, Dongdaemun-gu, Seoul 02455, Korea}

\author{Yuan Yin}
\email{yinyuan@kias.re.kr}
\affiliation{School of Physics, KIAS, 85 Hoegi-ro, Dongdaemun-gu, Seoul 02455, Korea}

\author{Bingrong Yu}
\email{bingrong.yu@cornell.edu}
\affiliation{Department of Physics, LEPP, Cornell University, Ithaca, NY 14853, USA}
%
%%%%%%%%%%%%%%%%%%%%%%%%%%%%%

%%%%%%%%%%%%%%%%%%%%%%%%%%%%%
\begin{abstract}
We construct a technically natural model in which thermal dark matter (DM) interacts with axion dark energy (DE) and produces an apparent late-time crossing of the phantom divide. A direct axion coupling to weakly-interacting massive particles (WIMPs) would ordinarily radiatively destabilize the ultralight axion potential. We avoid this issue through $N$ fermion species related by a cyclic $\mathbb Z_N$ symmetry, which projects the leading Coleman–Weinberg potential onto the exponentially suppressed $N$th harmonic. Although the microscopic theory preserves $\mathbb Z_N$, the axion-dependent WIMP masses generate unequal equilibrium abundances that freeze-out imprints on the cosmological relic state, thereby breaking the symmetry spontaneously. The resulting relic distribution retains a memory of the initial axion value and generates an unsuppressed finite-density potential that holds the field fixed at early times. As the WIMP density dilutes, the axion rolls  toward the minimum of its confining potential, transferring energy from DE to DM at late times. An observer assuming separately conserved components then infers an effective equation of state that crosses below $-1$, without ghosts or violation of the null-energy condition. We present an illustrative cosmological solution with a DESI-like phantom crossing and percent-level suppression of structure growth, and discuss the implications of the WIMP multiplicity and relic distribution for DM searches.
\end{abstract}
%%%%%%%%%%%%%%%%%%%%%%%%%%%%%

\maketitle
%\tableofcontents
\section{Introduction}\label{sec:intro}

``\textit{$\Lambda$ gets by with a little help from the WIMP.}'' In this work, we show that the relic density of weakly interacting massive particles can generate the finite-density potential required to trigger the late-time evolution of an axion dark-energy (DE) field. 
For more than two decades following the discovery of cosmic acceleration~\cite{SupernovaSearchTeam:1998fmf}, the cosmological constant $\Lambda$ has provided a remarkably successful description of the late-time Universe. Together with cold dark matter (DM), it forms the $\Lambda$CDM model, in which DE is identified with a constant vacuum energy density obeying the equation of state $w_\Lambda\equiv p_\Lambda/\rho_\Lambda=-1$. This minimal framework remains in excellent agreement with a broad range of observations, including Type-Ia supernovae (SNe), the cosmic microwave background (CMB), and the large-scale distribution of matter.

The latest baryon acoustic oscillation (BAO) measurements from the Dark Energy Spectroscopic Instrument (DESI), however, may be pointing beyond this simple picture~\cite{DESI:2024mwx,DESI:2025zgx}. When combined with SNe data~\cite{Brout:2022vxf,Rubin:2023jdq,DES:2024jxu} and CMB measurements~\cite{Planck:2018nkj,Planck:2019nip,Efstathiou:2019mdh,Carron:2022eyg,Rosenberg:2022sdy,ACT:2023kun,ACT:2023oei}, and interpreted within the Chevallier-Polarski-Linder (CPL) parametrization $w_{\rm DE}=w_0+w_a(1-a)$, with $a$ the scale factor of the Universe, the preferred region lies around $w_0>-1$, $w_a<0$, and $w_0+w_a<-1$. This behavior corresponds to an effective crossing of the phantom divide during the recent cosmological past. Its statistical significance and detailed redshift dependence remain sensitive to the choice of datasets and parametrization~\cite{Jiang:2024xnu,DESI:2025fii,Wolf:2025jlc,Gialamas:2025pwv,Toomey:2025xyo,Lee:2025pzo}, but the possibility provides strong motivation to explore theoretically consistent realizations of evolving DE.\\

A slowly evolving scalar field provides a simple realization of dynamical DE~\cite{Ratra:1987rm,Wetterich:1987fm}, as in quintessence~\cite{Caldwell:1997ii}. Obtaining $w_{\rm DE}<-1$ is more challenging~\cite{Caldwell:1999ew}. An isolated canonical scalar, and more generally any isolated component satisfying the null-energy condition (NEC), obeys $\rho+p\geq0$ and hence $w\geqslant-1$~\cite{Qiu:2007fd,Ludwick:2017tox,Moghtaderi:2025cns,Cai:2025mas,Caldwell:2025inn}. Dark-sector interactions offer a healthy alternative~\cite{Das:2005yj,Smith:2024ibv,Khoury:2025txd,Bedroya:2025fwh,Wang:2025znm,Guedezounme:2025wav,Chen:2025ywv,LaPenna:2026avs,Antusch:2026ldp,Delaunay:2026jto,Khoury:2026svx}: if DM exchanges energy with a canonical DE field, an observer assuming separately conserved components may infer $w_{\rm DE}<-1$, even though the underlying theory contains no ghosts and respects the NEC.\\

Among the numerous candidates for dynamical DE~\cite{Tsujikawa:2013fta},
axion-like fields~\cite{Frieman:1995pm, Kim:1998kx, Kim:1999dc, Liu:2025bss} are particularly well-motivated because they provide a technically natural explanation
for an ultralight scalar degree of freedom. In this framework, the shift
symmetry of the pseudo Nambu-Goldstone boson $\phi$, arising from
a spontaneously broken global Peccei-Quinn (PQ) symmetry, protects its ultralight mass from
radiative corrections. A mixed anomaly between the PQ symmetry and a confining gauge group $G_D$ can then generate the potential
\begin{equation}\label{eq:Vconf-intro}
    V(\phi)
    =
    \Lambda_0
    \left[
        1-\cos\left(\frac{\phi}{f}\right)
    \right] ,
\end{equation}
where $\Lambda_0$ is the topological susceptibility of $G_D$ and $f$ is the decay
constant of $\phi$. For an initial displacement of order \(f\), the field
remains frozen by Hubble friction until late times, when it begins to roll to its nearest minimum.
The observed DE density requires \(\Lambda_0\sim{\cal O}({\rm meV}^4)\). For the
field to roll at present, its mass must satisfy \(m_\phi\sim H_0\), with
\(H_0\sim{\cal O}(10^{-33}\,{\rm eV})\) the present-day Hubble parameter.
These two requirements imply \(f\sim \sqrt{\Lambda_0}/m_\phi\sim{\cal O}(10^{18}\,{\rm GeV})\), placing the decay constant close to the Planck scale, reflecting the well-known coincidence that the onset of cosmic acceleration occurs near the present Hubble scale.\\

We investigate whether such an axion can interact appreciably with thermal WIMPs. WIMP DM provides a predictive setting in which the relic abundance is determined by freeze-out~\cite{Lee:1977ua,Steigman:1984ac}, while the same interactions can be tested by direct detection~\cite{Goodman:1984dc}, indirect searches, and colliders; see Ref.~\cite{Cirelli:2024ssz} for a recent review. 

A direct nonderivative interaction between an ultralight scalar and such a massive particle would, however, radiatively destabilize the DE potential.
For example, the coupling $y\phi\bar\chi\chi$ to a fermionic WIMP $\chi$  generates a quadratically divergent correction at one-loop $\delta m_\phi^2\sim y^2\Lambda^2/(16\pi^2)$. Even for the low cutoff $\Lambda\sim4\pi m_\chi$, 
\begin{equation}
\frac{\delta m_\phi^2}{m_\phi^2}
\sim
10^{88}y^2 
\left(\frac{m_\chi}{100\,\GeV}\right)^2
\left(\frac{H_0}{m_\phi}\right)^2,
\end{equation}
so an appreciable DE--DM interaction requires an extraordinary fine-tuning to preserve the ultralight scalar mass.

The axion shift symmetry forbids such polynomial couplings, but it does not by itself solve the problem: if $\phi$ enters a fermion mass only through an overall phase, its dependence can be removed by a chiral field redefinition. A physical $\phi$ dependence therefore requires at least two mass contributions with different PQ transformation properties. Their interference explicitly breaks the continuous shift symmetry and induces a Coleman-Weinberg (CW) potential that is still generically much larger than~\cref{eq:Vconf-intro}.

We suppress this correction by introducing $N$ fermionic WIMPs related by a discrete $\mathbb{Z}_N$ symmetry acting through the cyclic transformation $\chi_k\to\chi_{k+1}$ and $\phi\to\phi+2\pi f/N$. The sum over the complete $\mathbb Z_N$ orbit projects the CW potential onto harmonics that are integer multiples of $N$, with leading contribution~\cite{Hook:2018jle,Brzeminski:2020uhm}
\begin{equation}
    V_{\rm CW}^{\chi}(\phi)
    \sim
    \epsilon^N\frac{m_\chi^4}{16\pi^2}
    \cos\left(\frac{N\phi}{f}\right),
    \label{eq:CW-intro}
\end{equation}
where $\epsilon\lesssim1$ controls the axion dependence of the individual WIMP masses. Choosing sufficiently large $N$ ensures $|V_{\rm CW}^{\chi}|\ll\Lambda_0$ without suppressing the microscopic axion--WIMP interaction. For $m_\chi=100\,\GeV$, for instance, $\epsilon=0.01\,(0.3)$ requires approximately $N\gtrsim25\,(100)$, with only a mild dependence on $m_\chi$.\\

Moreover, the SM interactions necessary to thermalize the WIMPs must respect also the $\mathbb Z_N$ structure. Sector-
dependent portal couplings would generate
mass corrections that spoil the cyclic cancellation, and induce
an unsuppressed axion potential. Selective thermalization is therefore incompatible with radiative stability, as is asymmetric reheating of replicated SM sectors, as discussed in~\cref{app:asymmetricReheating}.

The central observation of this work is that cosmological WIMP background must break the $\mathbb{Z}_N$ symmetry of the microscopic interactions. The relic DM energy density is
\begin{equation}
\rho_{\chi}(\phi)\equiv n_\chi\sum_{k=0}^{N-1} m_k(\phi)\eta_k\,,\quad \eta_k\equiv \frac{n_k}{n_\chi}\,,
\end{equation}
where $n_\chi\propto a^{-3}$ is the total number density. For a uniform distribution of the relic density across the $N$ sectors, $\eta_k=1/N$, the sum covers a complete $\mathbb Z_N$ orbit and the field dependence $\rho_\chi$ is suppressed by $\epsilon^N$, as in the CW vacuum potential. Efficient late-time energy transfer
therefore requires a nonuniform relic distribution.

Such a distribution arises naturally from thermal freeze-out. Inflation sets an approximately homogeneous axion value $\phi_i$, and hence a background-dependent WIMP spectrum, while the underlying Lagrangian still respects $\mathbb{Z}_N$. At high temperatures, the WIMPs are relativistic, their equilibrium populations are uniform, and their thermal contribution to the axion potential retains the cyclic suppression. 
Once the WIMPs become nonrelativistic, however, their different $\phi_i$-dependent masses produce unequal equilibrium densities that are imprinted in the relic fractions when inter-sector conversion processes decouple. The relic distribution thus retains a memory of the initial axion value, thereby transfering its spontaneous $\mathbb{Z}_N$ breaking to the WIMP background.

The resulting asymmetric cosmological state generates an unsuppressed finite-density potential for the axion. Since the sectors that were lightest at conversion decoupling are the most populated, the WIMP energy density is minimized when the instantaneous mass spectrum reproduces its freeze-out configuration. The potential therefore depends only on the relative displacement $\phi-\phi_i$ and is minimized at $\phi=\phi_i$.

At early times, the finite-density contribution holds the axion at $\phi_i$. As $n_\chi$ redshifts, this contribution weakens until the confinement potential takes over and the field rolls toward its nearest vacuum. Along this trajectory, the relic-weighted WIMP mass increases, transferring energy from DE to DM. An observer assuming separately conserved components consequently infers an effective DE equation of state that crosses below $-1$. We exhibit an illustrative cosmological solution with a DESI-like phantom crossing and a percent-level suppression of structure growth, and study how the multiplicity and relic distribution modify direct and indirect WIMP searches.\\

The paper is organized as follows. In~\cref{sec:Model}, we
introduce the model of axion DE interacting with $N$ copies of a fermionic WIMP, and its UV completion. In~\cref{sec:ZNfreezeout}, we compute the distribution of the WIMP DM relic abundance across the $N$ sectors. In~\cref{sec:potential}, we derive the axion potential in
the presence of this WIMP density. In~\cref{sec:phantomDE}, we study the resulting background evolution of the axion, paying particular attention to the  phantom crossing of the effective DE equation of state, and discuss its impact on the growth of matter perturbations. In ~\cref{sec:DMpheno}, we highlight the generic implications of the WIMP multiplicity on DM phenomenology. We conclude in~\cref{sec:conclusions}.

\section{Model of Axion--WIMP Interactions}
\label{sec:Model}
We consider a complex Peccei--Quinn (PQ) scalar
\begin{equation}
    \Phi = \frac{f+\rho}{\sqrt{2}}e^{i\phi/f}\,,
\end{equation}
whose angular component $\phi$ plays the role of the axion. To couple it to WIMP-like DM without destabilizing its potential, we introduce $N$ copies of a Dirac fermion, $\chi_k$ with $k=0,\dots,N-1$, all singlets under the confining gauge group introduced below. Their interactions are described by
\begin{align}
     \mathcal{L}_\chi\supset-\sum_{k=0}^{N-1}\left(M+y_\chi e^{2\pi i k/N}\Phi\right)\bar \chi_{Lk}\chi_{Rk} +{\rm h.c.}\,,
     \label{eq:Lag}
\end{align}
where the bare mass $M$ and the Yukawa coupling $y_\chi$ are universal across all sectors. The Lagrangian in~\cref{eq:Lag} is invariant under a discrete $\mathbb{Z}_N$ symmetry acting as a cyclic shift among the $N$ sectors,
\begin{equation}
    \chi_k\to \chi_{k+1}\,,\qquad
    \Phi\to \Phi\, e^{2\pi i/N}\,,
    \label{eq:ZNsym}
\end{equation}
with $\chi_{N-1}\to \chi_0$. This cyclic symmetry is the central ingredient protecting the axion potential from WIMP-induced radiative corrections.

Assuming ${\rm PQ}(\chi_{Lk})-{\rm PQ}(\chi_{Rk})={\rm PQ}(\Phi)=1$, the bare mass parameter explicitly breaks the continuous PQ symmetry while preserving its $\mathbb{Z}_N$ subgroup. The parameters $M$ and $y_\chi$ are in general complex, with one reparametrization-invariant phase $\theta_\chi\equiv {\rm arg}(M^{-1}y_\chi)$, associated with a $\mathbb{Z}_N$-symmetric chiral rotation of the $\chi_k$'s. After PQ breaking, the field-dependent complex mass parameters are
\begin{equation}
    m_k(\Theta)\equiv M+m_\chi \exp\left[i\left(\Theta+\frac{2\pi k}{N}+\theta_\chi\right)\right]\,,
\end{equation}
where $\Theta\equiv \phi/f$, and $m_\chi\equiv y_\chi f/\sqrt{2}$ and $M$ can be taken real and positive without loss of generality. 

To generate the vacuum potential responsible for DE, we couple the axion to a confining dark gauge group $G_D={\rm SU}(N_c)$ through
\begin{equation}
    \mathcal{L}_D\supset
    \frac{\alpha_D}{4\pi}
    \left(\frac{\phi}{f}+\theta\right)
    {\rm tr}\,G_{\mu\nu}\widetilde G^{\mu\nu}\,,
    \label{eq:LdarkQCD}
\end{equation}
where $\alpha_D$ is the dark gauge coupling, $G_{\mu\nu}$ is the corresponding field strength, $\widetilde G_{\mu\nu}\equiv \epsilon_{\mu\nu\rho\sigma}G^{\rho\sigma}/2$ its dual, and $\theta$ is the topological phase of the SU($N_c$) vacuum. Canonical kinetic terms are understood. Once the dark gauge theory confines, nonperturbative effects reduce the continuous shift symmetry to $\phi\to\phi+2\pi f$ and generate the $2\pi f$-periodic potential in~\cref{eq:Vconf-intro}, whose amplitude accounts for most of the DE density.
The phase $\theta$ can be removed from the topological coupling by a constant shift of $\phi$, leaving one physical relative phase $\bar\theta\equiv\theta-\theta_\chi$. As we show in~\cref{sec:ZNfreezeout}, this phase does not affect the displacement between the initial axion value and the extrema of the WIMP-induced finite-density potential.\\

The interaction in~\cref{eq:LdarkQCD} may arise, for instance, from a KSVZ-like completion~\cite{Kim:1979if, Shifman:1979if}. In its minimal realization, one introduces a heavy vector-like fermion $Q$ transforming in the fundamental representation of SU($N_c$), with PQ charges ${\rm PQ}(Q_L)-{\rm PQ}(Q_R)=1$. This chiral charge assignment forbids a bare Dirac mass for $Q$ while allowing the Yukawa interaction
\begin{equation}
    \mathcal{L}_{\rm KSVZ}\supset - y_Q\Phi \bar Q_L Q_R +{\rm h.c.}\,,
\end{equation}
where $y_Q$ is in general complex. After spontaneous PQ breaking, the field-dependent heavy-fermion mass is
\begin{equation}
    m_Q(\phi) =\frac{y_Q f}{\sqrt{2}}e^{i\phi/f}\,.
\end{equation}
Owing to the mixed PQ--SU($N_c)^2$ anomaly, removing its phase by a chiral rotation of $Q$ generates the interaction in~\cref{eq:LdarkQCD}, with $\theta\equiv\theta_G+\arg y_Q$, where $\theta_G$ is the microscopic topological angle.

At first sight, the anomalous interaction in~\cref{eq:LdarkQCD} appears incompatible with the $\mathbb{Z}_N$ symmetry of the WIMP sector. (The confining potential is $2\pi f$ periodic rather than $2\pi f/N$ periodic.) This difference arises because $\mathbb{Z}_N$ is a fundamental symmetry of the perturbative Lagrangian but has a mixed anomaly with SU($N_c$). In the KSVZ completion, it acts as
\begin{equation}
    \Phi\to e^{2\pi i/N}\Phi\,,\qquad
    \bar Q_L Q_R\to e^{-2\pi i/N}\bar Q_L Q_R\,,
\end{equation}
so that the KSVZ Yukawa interaction remains $\mathbb{Z}_N$ invariant, while the continuous PQ emerges as an accidental symmetry of the renormalizable KSVZ Lagrangian. The mixed $\mathbb Z_N$--SU$(N_c)^2$ anomaly violates the shorter shift $\phi\to\phi+2\pi f/N$ and, after confinement, generates the $2\pi f$-periodic potential in~\cref{eq:Vconf}.

Hence, the tiny dark topological susceptibility is the only spurion that generates lower harmonics forbidden by the perturbative $\mathbb Z_N$ structure. Any such contribution to the WIMP masses or their CW potential must contain an insertion of this nonperturbative spurion. The cyclic structure of~\cref{eq:Lag} is consequently preserved to excellent accuracy.\\

The amplitude of the axion-induced modulation of the WIMP masses is controlled by the ratio $\epsilon\equiv m_\chi/M$. An appreciable modulation therefore requires the bare mass and the contribution generated by PQ breaking to be comparable. This coincidence can arise naturally in UV realizations in which both contributions originate from a common symmetry-breaking scale. As an example, consider introducing an additional scalar $\Sigma$ with the leading interactions
\begin{equation}
  \mathcal{L}_\chi^{\rm UV}\supset
  -\sum_{k=0}^{N-1}
  \left(y_0+y_1e^{2\pi ik/N}\frac{\Phi}{\Lambda}\right)
  \Sigma\bar\chi_{Lk} \chi_{Rk}
  +{\rm h.c.}\,,
\end{equation}
where $\Lambda$ denotes the UV cutoff. These operators are invariant under an independent U(1)$_X$ symmetry with charge assignment $X(\Sigma)=X(\chi_{Lk})=1$ and $X(\Phi)=X(\chi_{Rk})=0$, while only the second operator preserves the PQ symmetry for ${\rm PQ}(\Sigma)=0$. After spontaneous breaking of U(1)$_X$, one recovers the effective interactions in~\cref{eq:Lag}, with
\begin{equation}
    M\equiv y_0\langle \Sigma\rangle\,,\qquad
    m_\chi=\frac{y_1}{\sqrt{2}}\langle \Sigma\rangle\frac{f}{\Lambda}\,,
\end{equation}
and hence
\begin{equation}
    \epsilon=\frac{y_1f}{\sqrt{2}y_0\Lambda}\,.
\end{equation}
For Yukawa couplings of comparable magnitude and a PQ-breaking scale not far below the cutoff, $f\lesssim\Lambda$, one naturally obtains $\epsilon\lesssim\mathcal{O}(1)$. Since no operator explicitly breaks U(1)$_X$, the fermion masses are independent of the phase of $\Sigma$. The corresponding Goldstone boson may be removed by gauging U(1)$_X$ in the UV, provided that the U(1)$_X^3$ and mixed U(1)$_X$-gravitational anomalies are cancelled by spectator fermions.

%%%%%%%%%%%%%%%%%%%%%%
\section{$\mathbb{Z}_N$ Breaking from Freeze-Out}\label{sec:ZNfreezeout}

At high temperatures, the $N$ fermion species are maintained in thermal equilibrium with the SM bath through a $\mathbb{Z}_N$-symmetric portal, for instance through universal electroweak interactions. Once they become nonrelativistic, their number densities obey a system of $N$ coupled Boltzmann equations~\cite{Griest:1990kh,Gondolo:1990dk}
\begin{align}
\dot n_k + 3H n_k =& -\langle \sigma v\rangle_{k\to {\rm SM}}\left(n_k^2-n_{k{\rm eq}}^2\right)\nonumber\\
&-\sum_{j\neq k}\langle\sigma v\rangle_{k\to j}\left(n_k^2-n_j^2\frac{n_{k{\rm eq}}^2}{n_{j{\rm eq}}^2}\right)\,,
\end{align}
where $n_{k{\rm eq}}$ is the equilibrium density. The first term on the right-hand side describes annihilations into SM particles, while the second accounts for inter-sector conversion reactions that redistribute the abundance among the $N$ sectors. In the nonrelativistic limit, 
\begin{equation}
    n_{k{\rm eq}}\simeq g_k\left(\frac{m_kT}{2\pi}\right)^{3/2}e^{-m_k/T}\,,
\end{equation}
where $g_k=4$ counts the two spin states of the fermion and antifermion, and assuming kinetic equilibrium with SM plasma 
\begin{equation}
n_k =e^{\mu_k/T} n_{k{\rm eq}}\,,
\end{equation}
where $\mu_k$ is an effective chemical potential. 

The fermion masses are not degenerate but exhibit a small mass difference coming from the sector-dependent phase. As shown in~\cref{app:finiteT}, the axion remains overdamped before and during freeze-out, and therefore fixed at its initial value $\phi_i$, and the fermion masses are
\begin{equation}
   m_k(\Theta_i)=M \sqrt{1+2\epsilon\cos\left(\Theta_i-\bar\theta+\frac{2\pi k}{N}\right)+\epsilon^2}\,. 
\end{equation}
For $\epsilon\lesssim 1$, the lightest state is in the sector $k=k_{\rm min}$ for which  $\Theta_i-\bar\theta+2\pi k/N$ lies closest to $\pi$ (modulo $2\pi$). More precisely, $\cos(\Theta_i-\bar\theta+2\pi k_{\rm min}/N)\equiv\cos(\pi+\delta)$, where the angle $\delta$ satisfies $|\delta|\leqslant\pi/N$. Conversely, for large $N$, the heaviest state is in the sector $k=k_{\rm max}$ where 
$k_{\rm max}\simeq k_{\rm min}+N/2$ (modulo $N$). As a result, at leading order in small $\epsilon$ and large $N$, all the states form a mass band of relative width $2\epsilon\cos\delta\simeq 2\epsilon(1-\delta^2/2)$.\\

The complete Boltzmann system is invariant under the combined
$\mathbb Z_N$ transformation in~\cref{eq:ZNsym}. A fixed initial axion
background $\phi_i$, however, selects a particular orientation in sector
space and thereby produces a nondegenerate WIMP spectrum. 
Although these mass splittings may be parametrically small, their
effect on the relative abundances can be enhanced by Boltzmann factors. Hence, the Boltzmann suppression of the heavier states produces a nonuniform distribution of the relic abundance across the $N$ sectors, where lighter states are more abundant. 
Therefore, freeze-out not only does fix the total DM abundance, but it also 
imprints the spontaneous $\mathbb{Z}_N$-breaking background selected by
$\phi_i$ onto the WIMP abundances. Once
inter-sector conversions decouple, this nonuniform distribution
persists even though the microscopic interactions remain
$\mathbb{Z}_N$ symmetric. As we show in~\cref{sec:potential}, this frozen
sector hierarchy is essential for the generation of the  finite-density
potential that controls the late-time axion dynamics.\\

Solving the full system of coupled Boltzmann equations is generally cumbersome. It simplifies considerably when the inter-sector conversion processes remain sufficiently rapid to maintain internal chemical equilibrium. In this regime, all species share a common chemical potential, $\mu_k=\mu$ for all $k$, and the fraction $\eta_k$ of the total number density within the $k$-th sector continues to track their corresponding equilibrium value~\cite{Griest:1990kh},
\begin{equation}
    \eta_k\equiv \frac{n_k}{n_\chi}\approx\frac{n_{k{\rm eq}}}{n_{\chi{\rm eq}}}\,,
\end{equation}
where $n_\chi\equiv \sum_k n_k$, as long as the conversion reactions remain efficient. 

Two distinct decoupling epochs must be distinguished. We denote by $T_f$
the freeze-out temperature of annihilations into SM states, which fixes
the total relic abundance, and by $T_d$ the temperature at which
inter-sector conversions cease to maintain internal chemical equilibrium.
Their ordering is model dependent. If $T_d>T_f$
conversions decouple first. The equilibrium fractions at $T_d$ then
provide the initial conditions for the subsequent sector-dependent
annihilation dynamics, and the final fractions must in general be
obtained from the full Boltzmann system.

We instead focus on $T_d\lesssim T_f$. In this regime, the total abundance freezes while the
sectors remain in internal chemical equilibrium, and the $N$ coupled Boltzmann equations collapse to the standard one of a single species~\cite{Griest:1990kh}
\begin{equation}\label{eq:Boltzmann}
    \dot n_\chi+3Hn_\chi=-\langle \sigma v\rangle_{\chi\to {\rm SM}}(n_\chi^2-n_{\chi {\rm eq}}^2)\,,
\end{equation}
where
\begin{equation}
    \langle\sigma v\rangle_{\chi\to{\rm SM}}\equiv \sum_k \eta_k^2\langle\sigma v\rangle_{k\to {\rm SM}}\,.
\end{equation}
After the total abundance has frozen, the conversion reactions may continue to redistribute the fixed number density among the sectors until $T=T_d$. 
In the  instantaneous decoupling approximation, and assuming that kinetic
equilibrium with the SM bath persists until $T_d$, the final fractions
are therefore
\begin{equation}\label{eq:etak}
    \eta_k^d=\frac{g_k}{g_{\chi}}\left(1+\Delta_k\right)^{3/2}e^{-x_d\Delta_k}\,,
\end{equation} 
where $x_d\equiv m_{k_{\rm min}}/T_d$, 
\begin{equation}
    g_{\chi}\equiv\sum_k g_k \left(1+\Delta_k\right)^{3/2}e^{-x_d\Delta_k}\,,
\end{equation}
and 
\begin{equation}
    \Delta_k\equiv \frac{m_k-m_{k_{\rm min}}}{m_{k_{\rm min}}}
\end{equation}
is the relative mass splitting between the $k$-th and lightest sectors.  At leading order in $\epsilon$,
$\Delta_k\propto
\epsilon$, 
so that the departure from a uniform distribution is controlled
parametrically by
$q_d\equiv\epsilon\left(x_d-3/2\right)$.  

\begin{figure}
    \centering
    \includegraphics[width=0.97\linewidth]{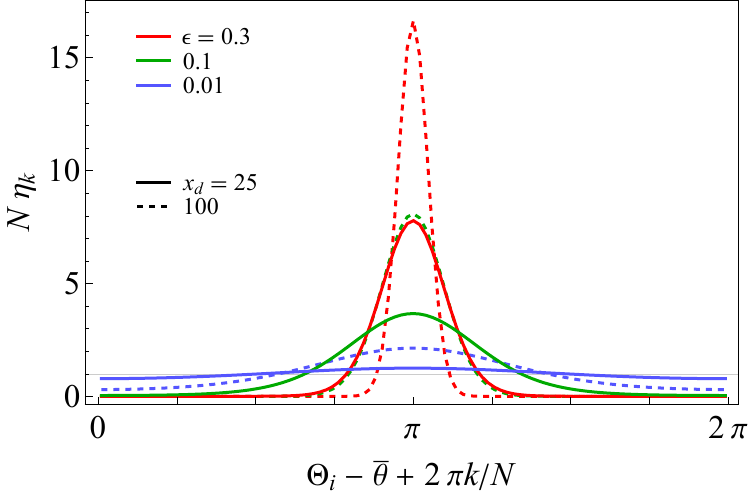}
    \caption{Relic abundance fraction $\eta_k$, normalized to the $\mathbb{Z}_N$-symmetric value $1/N$ indicated by the gray horizontal line, for $\epsilon=0.3$ (red), $0.1$ (green), and $0.01$ (blue), with conversion-decoupling temperature $x_d=x_f=25$ (solid) and $100$ (dashed). The lightest sectors, located near $\Theta_i-\bar\theta+2\pi k/N\sim \pi$, are preferentially populated. For $q_d\equiv\epsilon(x_d-3/2)\gtrsim 1$,  Boltzmann suppression of the heavier states makes the distribution increasingly peaked around the lightest sector.}
    \label{fig:relic-distrib}
\end{figure}

\Cref{fig:relic-distrib} shows the relic abundance distribution across the $N$ sectors for several values of $\epsilon$ and $x_d$. The lightest sectors, located near $\Theta_i-\bar\theta+2\pi k/N\sim \pi$, are preferentially populated, with this hierarchy becoming more pronounced for larger values of $q_d$ due to the stronger Boltzmann suppression of the heavier states. We defer the  implications of this nonuniform distribution for DM phenomenology to~\cref{sec:DMpheno}.

Inter-sector conversions can be mediated by the same universal interaction that thermalizes the WIMPs with the SM bath. In particular, if all copies couple to a common SM current or portal mediator, the processes $\chi_k\bar\chi_k\leftrightarrow\chi_j\bar\chi_j$ proceed through the exchange of a SM or portal state. Their rates are therefore controlled by the same portal couplings as annihilation
into SM states and may remain efficient through annihilation freeze-out. Whether this occurs
in practice depends on the mediator properties, the mass splittings, and the available final states. Below, we take $x_d\gtrsim x_f=25$ as the principal dynamical assumption of the freeze-out analysis, postponing its realization in specific portal models to future works.\\

%%%%%%%%%%%%%%%%%%%%%
\section{Axion Dark Energy potential}
\label{sec:potential}
%%%%%%%%%%%%%%%%%%%%%%

Here, we derive the effective potential governing the interplay between axion DE and finite-density WIMP DM. We first discuss the vacuum contributions and then the finite-density correction. Although only the vacuum potential enters the fluid equations in~\cref{sec:phantomDE}, the full effective potential clarifies the cosmological history of the model.

\subsection{Vacuum Potential}
We assume that the dark gauge sector is never appreciably populated after inflation and that its temperature remains below the confinement scale. Nevertheless, its zero-temperature topological susceptibility  generates a $2\pi f$-periodic potential for the axion~\cite{Witten:1998uka,Vicari:2008jw,Bonati:2016tvi,Ce:2016awn}, while potentially stable dark glueballs make a negligible contribution to the cosmic budget. 
At large $N_c$, the exact zero-temperature Yang-Mills vacuum energy exhibits a multibranched structure 
\begin{equation}
V_{\rm YM}(\Theta)=N_c^2\,\underset{\ell\in\mathbb{Z}}{\rm min}\,F\left(\frac{\Theta+2\pi\ell }{N_c}\right)\,,
\end{equation}
whose absolute minimum is at $\Theta=0$~\cite{Witten:1998uka}. For concreteness, we approximate its leading harmonic by a single cosine function~\cite{Bonati:2013tt}
\begin{equation}\label{eq:Vconf}
    V_D(\Theta)=\Lambda_D\left(1-\cos\Theta\right)\,,
\end{equation}
where $\Lambda_D$ is the topological susceptibility of the SU$(N_c)$ vacuum.\\ 

The explicit PQ breaking of the $\chi$ Lagrangian generates a one-loop potential for $\Theta$ through the CW mechanism~\cite{Coleman:1973jx} 
\begin{equation}\label{eq:CW}
    V_{\rm CW}(\Theta)=-\sum_{k=0}^{N-1}\frac{\big|m_k(\Theta)\big|^4}{16\pi^2}\left[\log\frac{|m_k(\Theta)\big|^2}{\mu^2}-\frac{3}{2}\right]\,,
\end{equation}
where 
\begin{align}
\big|m_k(\Theta)\big|^2 = M^2\left[1+\epsilon^2+2\epsilon \cos\left(\Theta-\bar\theta+\frac{2\pi k}{N}\right)\right]\,, \label{eq:mksq}
\end{align}
and $\epsilon\equiv m_\chi/M\geqslant0$. The physical $\Theta$ dependence arises from the interference between the shift-breaking mass $M$ and the shift-symmetric contribution $m_\chi$. Accordingly, when $m_\chi\to0$, or equivalently $\epsilon\to0$, the axion decouples from the fermion masses. Conversely, in the limit $M\to0$, the continuous shift symmetry is restored in the WIMP sector and the $\Theta$ dependence of the fermion mass can be removed by a suitable chiral field redefinition. In the following, we focus on the regime $\epsilon\lesssim1$, in which the CW potential admits a controlled expansion in powers of $\epsilon$.

Although~\cref{eq:CW} depends explicitly on the renormalization scale $\mu$, its field-dependent part is finite and independent of $\mu$ for $N>2$, resulting from the $\mathbb{Z}_N$ symmetry~\cite{Frieman:1991tu,Frieman:1995pm}. Indeed, the root-of-unity identities give $\sum_k\cos(x+2\pi k/N)=0$ and $\sum_k\cos^2(x+2\pi k/N)=N/2$ for $N>1$ and $N>2$, respectively. The $\mathbb{Z}_N$ symmetry also projects out all harmonics whose frequencies are not integer multiples of $N$. Consequently, for $\epsilon<1$, the first nontrivial $\phi$ dependence appears only at order $\epsilon^N$~\cite{Hook:2018jle}. Keeping the leading term in the small-$\epsilon$ expansion and subtracting a constant so that the potential vanishes at its minimum, one obtains for $N>2$
\begin{align}
    V_{\rm CW}(\Theta)\simeq (-1)^N\Lambda_N\cos\Big[N\left(\Theta-\bar\theta\right)\Big]\,,
\end{align}
where 
\begin{equation}
    \Lambda_N\equiv \frac{\epsilon^{N}M^4}{4\pi^2(N-1)(N-2)}\,.
\end{equation}
This potential contribution exhibits $N$ degenerate minima, located at $\Theta=\bar\theta+ 2k\pi /N$ for odd $N$, and $\Theta=\bar\theta+ (2k+1)\pi /N$ for even $N$, with $0\leqslant k\leqslant N-1$.\\

For $\Lambda_N\gg \Lambda_D$, matching the observed DE density requires $\Lambda_N\sim \meV^4$, while the scalar mass around a minimum is
\begin{equation}
m_\phi=\frac{N\sqrt{\Lambda_N}}{f}
\simeq\frac{M^2\epsilon^{N/2}}{2\pi f},
\end{equation}
where the second expression holds at large $N$. Requiring the field to evolve on cosmological time scales, $m_\phi\sim H_0\simeq2.13\,h\times10^{-33}\,\eV$, with $h\equiv H_0/(100\,{\rm km}\,{\rm s}^{-1}\,{\rm Mpc}^{-1})\simeq0.67$~\cite{Planck:2018vyg}, then gives 
$f\simeq 0.2(N/h)\MP$ 
where $\MP\simeq2.4\times10^{18}\,\GeV$ is the reduced Planck scale. In addition, for a weak-scale DM mass, reproducing the observed DE density requires the strong exponential suppression $\epsilon^N/N^2\sim4\times10^{-55}(100\,\GeV/M)^4$. As shown in~\cref{sec:phantomDE}, the DESI-preferred effective DE equation of state typically favors $\epsilon\sim0.01$--$0.3$, corresponding to $N\sim25$--$90$, with only a logarithmic sensitivity to the fermion mass scale. These requirements therefore point toward a decay constant significantly above $\MP$.

Interpreting such a trans-Planckian $f$ as a fundamental four-dimensional scale is potentially problematic from the EFT standpoint. Indeed, on general grounds~\cite{Banks:2010zn,Harlow:2018tng}, quantum-gravity (QG) is expected to explicitly break the global PQ and $\mathbb{Z}_N$ symmetries, inducing corrections to the axion potential which are no longer parametrically under control for $f\gg \MP$. (These symmetries cannot be gauged due to their mixed anomaly with $G_D$.) One possibility would be instead to interpret $f$ as an effective field range constructed from sub-Planckian microscopic scales and enhanced through an additional mechanism, such as clockwork~\cite{Choi:2015fiu,Kaplan:2015fuy}, alignment~\cite{Kim:2004rp}, or higher-dimensional gauge invariance~\cite{Arkani-Hamed:2003xts,Choi:2003wr}. 

Here, we adopt instead a more economical low-energy alternative and take $N$ sufficiently  large  such that the confining potential dominates over the CW contribution, $\Lambda_N\ll \Lambda_D$. In this regime, the axion mass  is  $m_\phi=\sqrt{\Lambda_D}/f$, so that requiring $\Lambda_D\sim {\rm meV}^4$ and $m_\phi\sim H_0$ implies
\begin{equation}
f\simeq0.2h^{-1}\MP\,. 
\end{equation}
The confining sector can therefore provide both the observed DE scale and a cosmologically light axion without requiring a trans-Planckian fundamental decay constant.

\subsection{Finite-Density Potential}
In addition to these vacuum contributions, the WIMP background in the early Universe generates additional contributions to the axion potential. As shown in \cref{app:finiteT}, the thermal contribution of uniformly distributed relativistic WIMPs is strongly suppressed by the underlying $\mathbb{Z}_N$ symmetry. However, once WIMPs become nonrelativistic and freeze out, the nonuniform distribution of their abundances generate an unsuppressed finite-density contribution.

For a dilute gas of nonrelativistic fermions, the leading finite-density contribution is
\begin{equation}
 \rho_\chi(\Theta)= m_\chi^{\rm eff}(\Theta)n_\chi\,,
\end{equation}
where $n_\chi$ is the total fermion number density and 
\begin{equation}\label{eq:mchieff}
    m_\chi^{\rm eff}(\Theta)\equiv\sum_{k=0}^{N-1} \eta_k^d|m_k(\Theta)|
\end{equation}
is the relic-weighted WIMP mass. Subtracting the field-independent contribution corresponding to the DM energy density $n_\chi m_\chi^{\rm eff}(\Theta_i)$ at conversion decoupling, the finite-density potential resulting from freeze-out is
\begin{equation}
    V_{\rm fd}(\Theta)=n_\chi \sum_{k=0}^{N-1} \eta_k^d\bigg[\left|m_k(\Theta)\right|-\left|m_k(\Theta_i)\right|\bigg]\,,
\end{equation}
Because the fermion masses differ across the $N$ sectors at freeze-out, the frozen fractions $\eta_k^d$ depart from the $\mathbb{Z}_N$-symmetric value $1/N$. Working at leading order in $\epsilon$ and in the large $N$ limit, the leading contribution is approximately (see~\cref{app:fdpotential} for details)
\begin{equation}
    V_{\rm fd}(\Theta)\simeq \sigma_{\rm fd}n_\chi\left[1-\cos\left(\Theta-\Theta_i\right)\right]+\mathcal{O}(\epsilon^2) \,,
\end{equation}
with 
\begin{equation}
    \sigma_{\rm fd}\simeq\epsilon M  \frac{I_1(q_d)}{I_0(q_d)}\,,
\end{equation}
where $I_0$ and $I_1$ are modified Bessel functions of the first kind and $q_d= \epsilon (x_d-3/2)$. The finite-density contribution is minimized at $\Theta=\Theta_i$.  Indeed, the sectors that were lightest at freeze-out acquire the largest relic fractions, and the total energy is subsequently minimized when the instantaneous mass spectrum realigns with the frozen abundance, which occurs precisely at the initial field value. As a result, the finite-density potential depends only on the relative displacement $\Theta-\Theta_i$, and is independent of the $\bar\theta$ phase.

Since $q_d>0$, $\sigma_{\rm fd}$ is strictly positive. For $q_d\ll 1$, corresponding to a mildly nonuniform distribution in which all $N$ sectors remain appreciably populated, $I_1(q_d)/I_0(q_d)\simeq q_d/2$ and therefore $\sigma_{\rm fd}/M\simeq \epsilon^2(x_d-3/2)/2$. In the opposite limit ($q_d\gg 1$), while keeping $\epsilon q_d \lesssim1$, $I_1(q_d)/I_0(q_d)\to 1$ so that  $\sigma_{\rm fd}/M \simeq \epsilon$. In this regime, the relic abundance is strongly concentrated in the sectors that were lightest at freeze-out, and the amplitude of the finite-density potential saturates, becoming independent of the conversion-decoupling temperature.\\

To summarize, in the regime of interest, $\Lambda_N\ll\Lambda_D$, where the $\mathbb{Z}_N$-symmetric CW contribution is subleading, the effective DE potential can  be written as
\begin{align}\label{eq:VDE}
    V_{\rm eff}(\Theta)=\,&V_0+\Lambda_D\left(1-\cos\Theta\right)\nonumber\\
    &+\sigma_{\rm fd}n_\chi\left[1-\cos\left(\Theta-\Theta_i\right)\right]\,,
\end{align}
where $V_0$ denotes a field-independent contribution whose value is determined by whatever mechanism ultimately addresses the cosmological-constant problem~\cite{Weinberg:1988cp}. 

\begin{figure}
    \centering
    \includegraphics[width=0.97\linewidth]{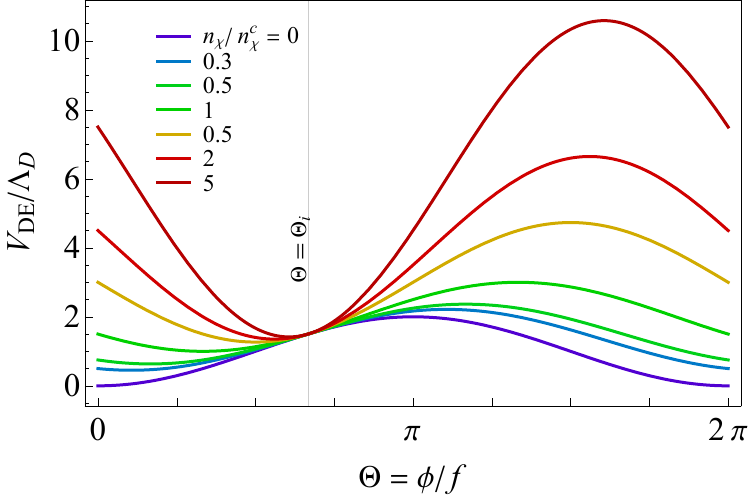}
    \caption{Effective DE potential $V_{\rm DE}$ for the parameters $V_0=0$, $\epsilon=1/10$, $x_d=x_f=25$, $\Theta_i=2\pi/3$, and several values of the DM density $n_\chi$. As $n_\chi$ increases, the minimum shifts from the vacuum minimum at $\Theta$ toward the density-induced minimum at $\Theta=\Theta_i$. The vacuum and finite-density contributions become comparable near $n_\chi=n_\chi^c\equiv \Lambda_D/\sigma_{\rm fd}$.}
    \label{fig:potential}
\end{figure}

\Cref{fig:potential} shows the evolution of the effective DE potential for different values of the DM density around the critical value $n_\chi^c\equiv \Lambda_D/\sigma_{\rm fd}$. For $n_\chi\gg n_\chi^c$, the finite-density potential dominates, trapping the field to its initial value $\Theta_i$. When $n_\chi\lesssim n_\chi^c$, the confining potential starts to dominate and the minimum of the effective potential moves toward the vacuum at $\Theta=0$ (modulo $2\pi$).

\section{Apparent Phantom Dark Energy}
\label{sec:phantomDE}

We now study the cosmological evolution of the axion $\phi$ as the field responsible for late-time DE, and identify the parameter space leading to a crossing of the phantom divide similar to DESI reconstruction. 
Because the effective WIMP mass depends on $\phi$, this evolution induces an exchange of energy between DE and DM, causing the energy density of the latter to deviate from the standard  scaling as $a^{-3}$. 
An observer interpreting the resulting cosmological evolution within a framework of noninteracting DE would therefore infer an effective equation of state~\cite{Das:2005yj,Khoury:2025txd},
\begin{align}\label{eq:wDE}
w_{\rm DE}^{\rm eff}=w_\phi\left[1+\frac{\rho_{\rm DM}}{\rho_\phi}\left[A(\phi)-1\right]\right]^{-1},
\end{align}
where $w_\phi\equiv p_\phi/\rho_\phi$, with $p_\phi\equiv \dot{\phi}^{ 2}/2-V$ and $\rho_\phi\equiv \dot{\phi}^{ 2}/2+V$ denoting the pressure and energy density of $\phi$, respectively, $V$ being the vacuum axion potential. Here $\rho_{\rm DM}\equiv m_\chi^{\rm eff}(\phi_0/f) n_\chi\propto a^{-3}$, with $\phi_0$ the present value of the axion field, is a reference DM energy density with the standard redshift scaling of pressureless matter, and  
\begin{equation}
    A(\phi)\equiv \frac{m_\chi^{\rm eff}(\phi/f)}{m_\chi^{\rm eff}(\phi_0/f)}\,.
\end{equation}
Therefore, phantom crossing ($w_{\rm DE}^{\rm eff}<-1$) is obtained for $A(\phi)<1$, that is, when the effective WIMP mass \emph{increases} at late times.

The time variation of the effective WIMP mass in~\cref{eq:mchieff} is given approximately by
\begin{equation}
\dot m_\chi^{\rm eff}(\Theta)
\simeq \sigma_{\rm fd}\dot\Theta\sin(\Theta-\Theta_i)\,.
\end{equation}
Once released from $\Theta_i$, the field rolls toward the nearest vacuum: $\Theta=0$ for $0<\Theta_i<\pi$ and $\Theta=2\pi$ for $\pi<\Theta_i<2\pi$. In the first case, both $\dot\Theta$ and $\sin(\Theta-\Theta_i)$ are negative, whereas in the second they are both positive. Hence, since $\sigma_{\rm fd}>0$, $\dot m_\chi^{\rm eff}>0$ along either trajectory, and the effective WIMP mass increases monotonically, as required by phantom crossing.\\

\subsection{Background evolution}
The origin of this effective equation of state can be understood from the coupled evolution of the DM and DE components. As a result of the energy transfer with $\phi$, the DM energy density obeys the modified continuity equation, 
\begin{equation}\label{eq:continuityeq}
\rho_\chi'+3\mathcal{H}\rho_\chi- \phi'\beta(\phi)\rho_\chi=0\,,
\end{equation}
where primes denote derivatives with respect to conformal time, $\mathcal{H}\equiv a'/a$ is the conformal Hubble parameter, and $\beta(\phi)\equiv {\rm d}\log m_\chi^{\rm eff}/{\rm d}\phi$.  
The evolution of the axion field is governed by the Klein-Gordon equation,
\begin{equation}\label{eq:KGphi}
\phi''+2\mathcal{H}\phi'+a^2\left[\frac{{\rm d}V}{{\rm d}\phi}+\beta(\phi)\rho_\chi\right]=0\,,
\end{equation}
where the second term inside the brackets arises from the 
$\phi$-dependence of the effective WIMP mass and represents the backreaction of the DM density on the axion dynamics.\\

The axion field may be initially displaced from the minimum of its potential on cosmological scales if the PQ symmetry is broken before inflation. Inflation then stretches a single Hubble patch, characterized by a random initial value of the field $\phi_i$, to encompass the entire observable Universe, resulting in an effectively homogeneous axion field. To determine its subsequent evolution, we solve the coupled system in~\cref{eq:continuityeq} and \cref{eq:KGphi} starting deep in the radiation era at $z_i=10^7$, with initial conditions  $\phi(z_i)=\phi_i\sim \mathcal{O}(f)$ and  $\phi'(z_i)=0$ for the axion field. 

The initial radiation density is fixed by the observed CMB photon temperature~\cite{Fixsen:2009ug}, assuming three relativistic neutrino species and neglecting the small contribution of neutrino masses to the late-time energy budget. Furthermore, we adopt the $\Lambda$CDM values~\cite{Planck:2018vyg} for the baryon abundance and Hubble at $z_i$, while the initial WIMP abundance is fixed through the flatness condition. As a result, the expansion history at early times remains essentially identical to that of the standard $\Lambda$CDM cosmology, with deviations appearing only when the axion field begins to evolve dynamically at late times.\\

Four parameters control the axion dynamics: the  vacuum-energy offset $V_0$, the overall scale of the vacuum potential $\Lambda_D$, the WIMP mass-interference ratio $\epsilon$, and the axion decay constant $f$. Before freeze-out, the $\mathbb{Z}_N$ symmetry suppresses the finite-temperature contribution to the axion potential. The axion mass is therefore set by the confining potential, $m_\phi=\sqrt{\Lambda_D}/f\sim H_0$,
and Hubble friction keeps the axion field  frozen at $\phi_i$. After freeze-out, the nonuniform distribution of the WIMP relic abundance among the $N$ sectors generates a finite-density potential, giving the axion an effective mass
\begin{equation}
m_{\phi}^2\big|_{\rm fd}=\frac{\sigma_{\rm fd}n_\chi}{f^2}\simeq \frac{\epsilon M }{f^2}n_\chi\,,
\end{equation}
where the second relation holds for $\epsilon (x_d-3/2)\gtrsim1$. Despite this large finite-density axion mass, which could exceed the Hubble parameter at early times, the same finite-density potential traps the axion at $\phi_i$. 

As the Universe expands and the WIMP density redshifts as $n_\chi\propto a^{-3}$, the finite-density contribution progressively weakens. Once  $n_\chi\leqslant\Lambda_D/\sigma_{\rm fd}$, the vacuum potential becomes dominant and the axion mass asymptotes to the constant value,
\begin{equation}
m_\phi^2\simeq\frac{\Lambda_D}{f^2}\,. 
\end{equation}
At this transition, the energy density stored in the axion field relative to that of the WIMP is $\rho_\phi/\rho_\chi\simeq \sigma_{\rm fd}/M\left(1-\cos\Theta_i+V_0/\Lambda_D\right)\lesssim 1$, so the Universe remains matter dominated. The weakening of the finite-density potential then releases the axion from the attractor, allowing it to roll toward the true vacuum at $\phi=0$ (modulo $2\pi$), provided $f\lesssim\sqrt{3\epsilon}\,M_{\rm P}$. The redshift of this transition is set primarily by $\Lambda_D$ and $\epsilon$, and is approximately
\begin{equation}
1+z\simeq 1.6\left(\frac{\epsilon}{0.1}\right)^{-1/3}\left(\frac{\Lambda_D}{H_0^2M_{\rm P}^2}\right)^{1/3}\,,
\end{equation}
where $\sqrt{H_0M_{\rm P}}\simeq 1.86\,{\rm meV}$. If the axion is released close to the maximum of its vacuum potential, $\Theta_i\simeq\pi$, its subsequent evolution—and hence the effective phantom crossing—can be significantly delayed, as the field initially takes time to build up velocity.
\begin{figure}
    \centering
\includegraphics[width=1\linewidth]{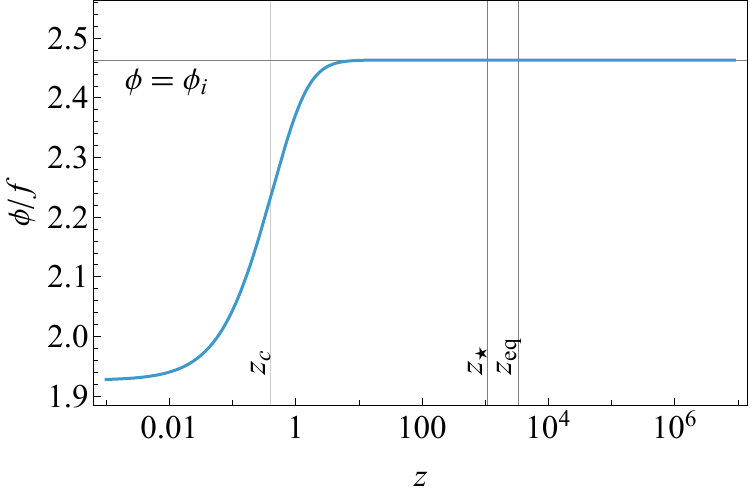}
    \caption{Evolution of the axion field $\phi$ as a function of redshift for a benchmark set of parameters, obtained by numerically solving the coupled DM--DE evolution equations. The field remains trapped to its initial value $\phi_i$ due to  the finite-density potential, then slowly rolls toward the vacuum minimum at $\phi=0$ once the finite-density contribution becomes subdominant. The vertical lines at $z_{\rm eq}\approx3381$, $z_\star\approx1090$, and $z_c\approx0.40$ denote the epochs of matter--radiation equality, recombination, and the effective phantom crossing predicted by the benchmark solution, respectively.}
    \label{fig:axionsolution}
\end{figure}

In~\cref{fig:axionsolution}, we present the numerical solution for the axion field evolution for a representative benchmark choice of the model parameters:
\begin{eqnarray}
V_0=\ 0.784H_0^2M_{\rm P}^2\,,\quad& \Lambda_D=0.784H_0^2M_{\rm P}^2\,,\quad&
     \epsilon=0.204\,,\quad \nonumber\\
    f=\ 0.327M_{\rm P}\,,\quad &\phi_i=0.784\pi f \,.
\end{eqnarray}
The benchmark point is chosen such that the angular acoustic scale $\theta_\star$ agrees with the Planck measurement~\cite{Planck:2018vyg} within one standard deviation.

\begin{figure}[t]
\centering
\includegraphics[width=\linewidth]{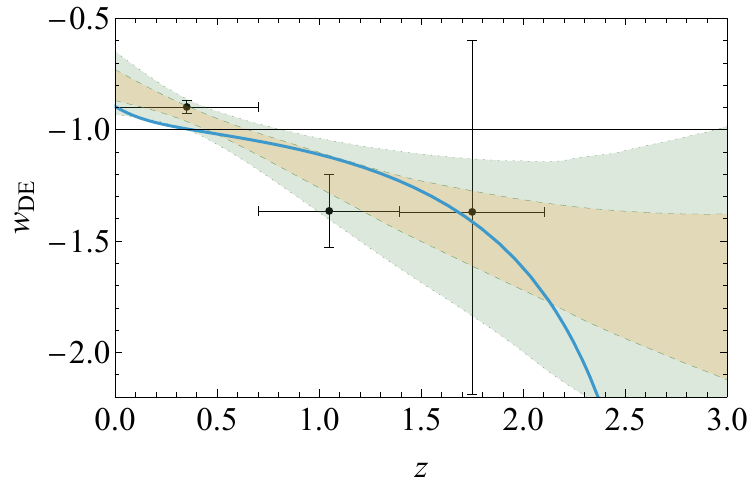}
\caption{Effective DE equation of state as a function of redshift for the benchmark solution, obtained by numerically solving the coupled DM--DE evolution equations (blue line). Black points with error bars show the binned DESI DR2 reconstruction~\cite{DESI:2024mwx}. The dark and light shaded bands indicate the $68\%$ and $95\%$ confidence regions, respectively, obtained from Gaussian-process reconstructions of $w_{\rm DE}(z)$ using the combined DESI, CMB, and Union3 datasets~\cite{DESI:2025fii}.}
\label{fig:wDEplot}
\end{figure}

The corresponding effective DE equation of state, defined in~\cref{eq:wDE}, is shown in~\Cref{fig:wDEplot}. As the field rolls toward the vacuum minimum at late times, the effective WIMP mass continuously increases. 
Consequently, the DM energy density redshifts more slowly than the standard $a^{-3}$ scaling, leading to an effective crossing of the phantom divide and qualitatively reproducing the trend suggested by the DESI reconstruction. For the benchmark solution, the crossing occurs at $z_c\simeq0.40$.

Departures from $\Lambda$CDM can also be characterized through the $Om$ diagnostic 
\begin{equation}
Om(z)\equiv\frac{h^2(z)-1}{(1+z)^3-1}\,,
\end{equation}
and the deceleration parameter,
\begin{equation}
    q(z)\equiv\frac{{\rm d}\log h}{{\rm d}\log(1+z)}-1\,.
\end{equation}
Both quantities depend only on the expansion history through the normalized Hubble parameter $h(z)\equiv H(z)/H_0$. In $\Lambda$CDM, $Om(z)=\Omega_m$ is constant, whereas $q(z)$ tracks the transition from decelerated to accelerated expansion.
\begin{figure}
\centering
\includegraphics[width=\linewidth]{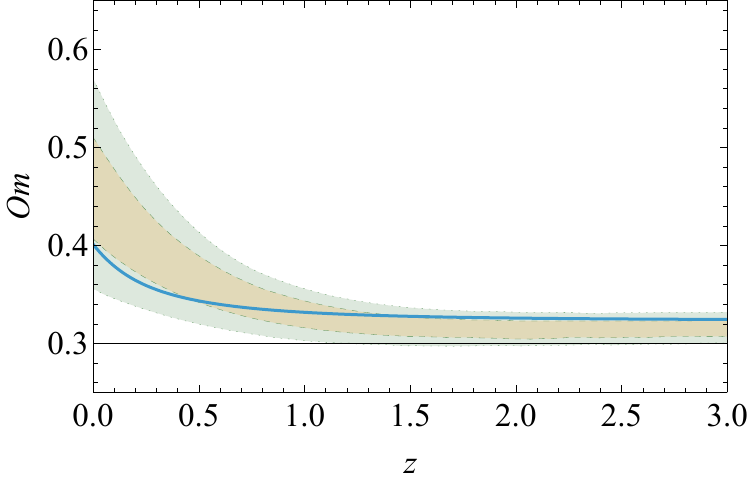}
\includegraphics[width=\linewidth]{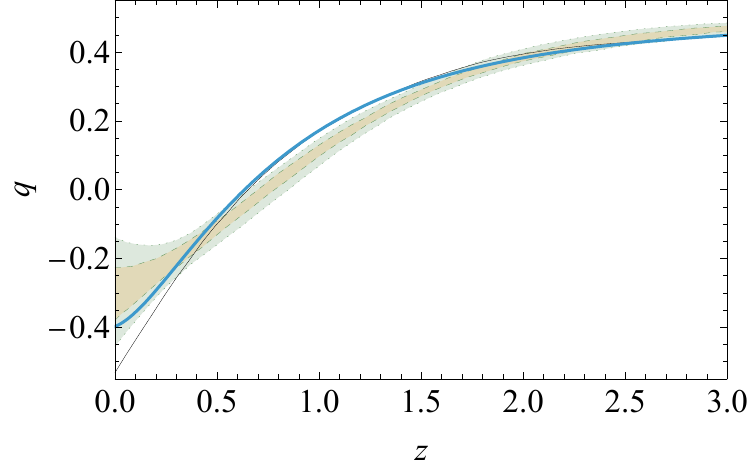}
\caption{Background diagnostics for the benchmark solution (blue lines). Top: $Om(z)$ diagnostic, which is constant and equal to $\Omega_m=0.302$ in $\Lambda$CDM. Bottom: deceleration parameter $q(z)$. The dark and light shaded bands indicate the $68\%$ and $95\%$ confidence regions, respectively, inferred from Gaussian-process reconstructions of $w_{\rm DE}(z)$ using the combined DESI, CMB, and Union3 datasets~\cite{DESI:2025fii}.}
\label{fig:de-diagnostics}
\end{figure}
As shown in~\cref{fig:de-diagnostics}, the late-time axion evolution produces correlated departures of $Om(z)$ and $q(z)$ from their $\Lambda$CDM predictions. In particular, the benchmark exhibits a recent weakening of cosmic acceleration, reflected in the upward evolution of $q(z)$ at the present epoch, as well as a 10\% increase of the total DM energy density relative to the $\Lambda$CDM value towards the end of the matter-dominated era, consistently with the DESI reconstruction of the expansion history~\cite{DESI:2025fii}.

\subsection{Linear Perturbations}
The coupling between the axion and WIMPs also modifies the evolution of cosmological perturbations. Since the WIMP mass depends on the local value of the axion field, spatial fluctuations of $\phi$ induce fluctuations in the DM mass, giving rise to an effective fifth-force interaction within the dark sector~\cite{Archidiacono:2022iuu,Bottaro:2023wkd,Bottaro:2024pcb}. Conversely, perturbations in the WIMP density modify the finite-density contribution to the axion potential, thereby sourcing axion fluctuations. Consequently, as shown in~\cref{app:linpert}, the linear perturbations of the DM and DE sectors become dynamically coupled, leading to departures from the standard $\Lambda$CDM growth history. 

The impact of the DM--DE interaction on structure formation can be quantified through $f\sigma_8(z)$, where $f\equiv {\rm d}\ln D/{\rm d}\ln a$ is the linear growth rate, with $D(a)$ the linear growth factor of matter perturbations, and $\sigma_8(z)$ is the root-mean-square amplitude of matter fluctuations smoothed over spheres of radius $8\,h^{-1}\,\mathrm{Mpc}$. The quantity $f\sigma_8$, which is directly constrained by redshift-space distortions (RSD)~\cite{Beutler:2012px,Howlett:2014opa,Blake:2013nif,WiggleZ:2012sek,Pezzotta:2016gbo,Okumura:2015lvp} and peculiar velocity surveys (PVS)~\cite{Said:2020epb,Boruah:2019icj,Carrick:2015xza,Huterer:2016uyq,Turner:2022mla}, provides a particularly robust probe of departures from the standard $\Lambda$CDM growth history. 

\begin{figure}
    \centering
\includegraphics[width=1\linewidth]{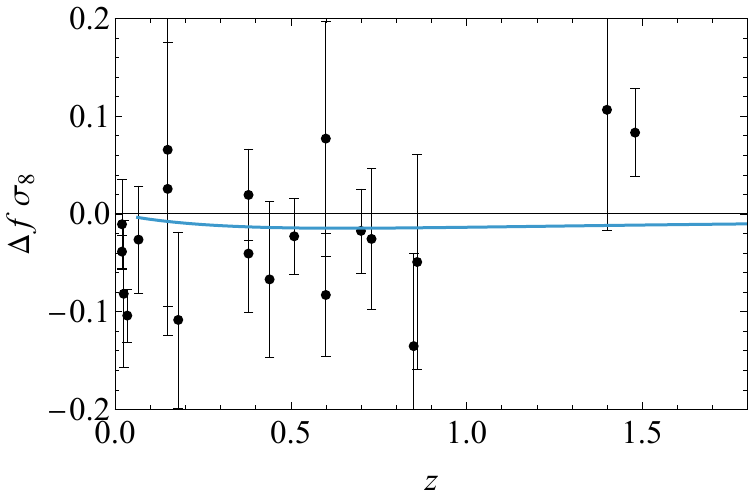}
    \caption{Evolution of the linear growth observable $\Delta f\sigma_8(z)\equiv f\sigma_8(z)-f\sigma_8^{\Lambda{\rm CDM}}(z)$  for a representative benchmark choice of the axion DE parameters (blue line), obtained by solving the coupled background and linear perturbation equations,  relative to the prediction of the $\Lambda$CDM model~\cite{Euclid:2025bxg}. The black points with error bars correspond to current $f\sigma_8$ determinations from redshift-space distortion measurements~\cite{Beutler:2012px,Howlett:2014opa,Blake:2013nif,WiggleZ:2012sek,Pezzotta:2016gbo,Okumura:2015lvp} and peculiar velocity surveys~\cite{Said:2020epb,Boruah:2019icj,Carrick:2015xza,Huterer:2016uyq,Turner:2022mla}, after subtracting the same reference $\Lambda$CDM prediction. The DE-DM interaction induced by the axion-dependent WIMP mass typically suppresses the growth of matter perturbations relative to $\Lambda$CDM by a few percent over the redshift range probed by current observations, while remaining compatible with existing data.}
    \label{fig:growth}
\end{figure}

In~\cref{fig:growth}, we present the shift 
\begin{equation}
    \Delta f\sigma_8(z)\equiv f\sigma_8(z)-f\sigma_8^{\Lambda{\rm CDM}}(z)
\end{equation}
in the growth observable $f\sigma_8(z)$ for the representative benchmark choice of the axion DE parameters discussed in~\cref{sec:phantomDE}, relative to the $\Lambda$CDM prediction~\cite{Euclid:2025bxg}. We find that the DM--DE interaction typically suppresses the growth of matter perturbations at the percent level relative to $\Lambda$CDM over the redshift range probed by current observations.
This suppression is not caused by the scalar fifth force, which tends to enhance DM clustering on scales below the Compton wavelength of the axion field~\cite{Archidiacono:2022iuu}. Instead, it results from the combined effect of the modified Hubble flow due to the time
dependence of the DM mass and the additional velocity drag
\(\beta\bar\phi'\theta_\chi\) during the epoch in which the effective
phantom behavior is generated.

%%%%%%

\section{Dark Matter Phenomenology}
\label{sec:DMpheno}
The multiplicity of the WIMP sector can substantially modify DM phenomenology. Assuming that conversion processes maintain internal chemical equilibrium throughout freeze-out, the total number density $n_\chi=\sum_k n_k$ obeys the standard Boltzmann equation in~\cref{eq:Boltzmann}, governed by the effective annihilation cross section $\langle\sigma v\rangle_{\chi\to{\rm SM}}=\sum_k \eta_k^2\langle\sigma v\rangle_{k\to {\rm SM}}$.  Since the WIMPs are already nonrelativistic at chemical decoupling, their contribution to the radiation density and hence to the Hubble expansion rate is Boltzmann suppressed. Reproducing the observed relic abundance therefore requires $\langle\sigma v\rangle_{\chi\to {\rm SM}}=\langle \sigma v\rangle_{\rm th}$, where $\langle \sigma v\rangle_{\rm th}\simeq 3\times 10^{-26}\,{\rm cm}^3/{\rm s}$ is the canonical thermal cross section for a single WIMP species~\cite{Gondolo:1990dk}. The phenomenological consequences depend on the effective number of sectors populated at freeze-out, 
\begin{equation}
N_f\equiv\left[\sum_k(\eta_k^f)^2\right]^{-1}\,,
\end{equation} where $\eta_k^f$ is the fraction of relic density in the $k$-sector at the time of annihilation decoupling, obtained from $\eta_k^d$ in~\cref{eq:etak} by replacing $x_d$ by $x_f\equiv m_{k_{\rm min}}/T_f$.  
For approximately equal microscopic annihilation cross sections, one then finds
\begin{equation}\label{eq:sigmaboost}
\langle\sigma v\rangle_{k\to{\rm SM}}
\simeq
N_f
\langle\sigma v\rangle_{\rm th}\,.
\end{equation}

When the relic distribution is strongly localized in the lightest sector, $N_f\simeq 1$,  the system effectively behaves as a single WIMP species, and the standard phenomenological expectations are approximately recovered.  The condition $q_f\equiv\epsilon(x_f-3/2)\gtrsim1$ marks the onset of a significantly nonuniform sector distribution. In the large-$N$ limit, 
\begin{equation}
\eta_k^f\simeq \frac{1}{NI_0(q_f)}\exp\left[-q_f\cos\left(\Theta_i-\bar\theta+\frac{2\pi k}{N}\right)\right]\,,
\end{equation}
which is strongly peaked around the lightest sector at $\Theta_i-\bar\theta+2\pi k_{\rm min}/N=\pi+\delta$ with $|\delta|\leqslant \pi/N$. Expanding the cosine around $\pi$, $\eta_k^f\propto \exp(-q_f\delta^2/2)$, implying that the  
the populated region has an angular width of order $1/\sqrt{q_f}$ and contains approximately $N/(2\pi\sqrt{q_f})$ sectors. Hence, the effective single-sector behavior requires $q_f\gtrsim N^2/2\pi$. For $1\ll q_f\ll N^2/2\pi$, the effective number of sectors populated at freeze-out is approximately $N_f\simeq N/\sqrt{\pi q_f}$.

In the opposite regime, $q_f\lesssim1$, the relic abundance is distributed approximately uniformly among all sectors,
\begin{equation}
\eta_k^f\simeq\frac{1}{N}
\left[
1-q_f\cos\left(\Theta_i-\bar\theta+\frac{2\pi k}{N}\right)
+\mathcal O(q_f^2)
\right]\,,
\end{equation}
so that $N_f\simeq N(1-q_f^2/2)$. The annihilation cross section of each individual species must then be enhanced by approximately a factor $N$ relative to the canonical thermal value.

The implications of the enhancement in~\cref{eq:sigmaboost} depend on the microscopic nature of the WIMP. For electroweak WIMPs, such as those considered in minimal-DM models~\cite{Cirelli:2005uq,Cirelli:2007xd,Cirelli:2009uv}, the relevant interactions are fixed by the SM gauge couplings. Assuming the parametric scaling $\langle\sigma v\rangle_{k\to{\rm SM}}\propto m_k^{-2}$, as in the case of $m_k\gg m_{\rm SM}$, implies a WIMP mass smaller by approximately a factor $\sqrt{N_f}$ than in the corresponding single-species model. Alternatively, for electroweak-singlet WIMPs coupled to the SM through a new portal coupling $g_{\rm NP}$, with $\langle\sigma v\rangle_{k\to{\rm SM}}\propto g_{\rm NP}^4$, the correct relic abundance may instead be obtained at fixed mass by increasing the portal coupling by a factor $N_f^{1/4}$.\\

While the relic-density requirement is controlled by the equilibrium sector distribution at annihilation freeze-out, present-day signals depend on the final distribution established at conversion decoupling. We therefore define
\begin{equation}
N_d\equiv \left[\sum_k (\eta_k^d)^2\right]^{-1}\,,
\end{equation}
in analogy with the effective multiplicity $N_f$ at annihilation freeze-out. Indirect-detection (ID) signals depend quadratically on the present-day component densities. Since only WIMPs belonging to the same sector can annihilate into SM particles, the total ID rate scales as 
\begin{equation}
    R_{\rm ID}\propto n_\chi^2\sum_k\eta_k^2\langle\sigma v\rangle_{k\to {\rm SM}}\simeq n_\chi^2\frac{N_f}{N_d}\langle\sigma v \rangle_{\rm th}\,,
\end{equation}
where we have assumed approximately sector-independent annihilation cross sections and used the relic-density requirement in~\cref{eq:sigmaboost}. If conversions decouple shortly after annihilation freeze-out, the sector distribution changes little, so that $N_d\simeq N_f$, and the standard indirect signal of thermal WIMPs is approximately recovered. If instead conversions remain efficient down to lower temperatures, the relic abundance becomes increasingly concentrated in the lightest sectors, yielding $N_d<N_f$ and enhancing the ID signal by a factor $N_f/N_d$ relative to the canonical thermal prediction. The maximal enhancement is $N_f$, reached when the relic abundance is entirely localized in a single sector ($N_d\simeq1$), and can therefore approach $N$ for a nearly uniform freeze-out distribution ($N_f\simeq N$).

For direct detection (DD), the total scattering rate is only linear in the density of each component, 
\begin{equation}
R_{\rm DD}\propto n_\chi\sum_k \eta_k\sigma_{k{\rm SM}}\,,    
\end{equation}
where $\sigma_{k{\rm SM}}$ is the scattering cross section of species $\chi_k$ off an SM target and $\eta_k$ denotes its present-day fraction, fixed at conversion decoupling. If the populated sectors have approximately equal scattering cross sections, $\sigma_{k{\rm SM}}\simeq\sigma_{\chi{\rm SM}}$, then $R_{\rm DD}\simeq n_\chi \sigma_{\chi{\rm SM}}$, so distributing the relic abundance among several sectors does not by itself suppress the total DD rate. The multiplicity nevertheless affects the rate indirectly through the microscopic parameters required by the relic-density condition. In particular, if the annihilation enhancement in~\cref{eq:sigmaboost} is obtained by increasing a portal coupling, and the annihilation and scattering cross sections have the same coupling dependence, the microscopic DD cross section, and hence the total DD rate at fixed WIMP mass, is enhanced by approximately a factor $N_f$ relative to the corresponding single-species model.

\section{Conclusions}
\label{sec:conclusions}

We have presented a radiatively stable framework in which an ultralight axion responsible for DE interacts appreciably with thermal WIMP DM. The central ingredient is a fundamental cyclic $\mathbb Z_N$ symmetry acting on $N$ fermionic WIMPs. Summing over the complete $\mathbb Z_N$ orbit suppresses the WIMP-induced CW potential, thereby protecting the axion from weak-scale radiative corrections.

Although the microscopic theory is $\mathbb Z_N$ symmetric, its cosmological state need not be. Once the WIMPs become nonrelativistic, their axion-dependent mass splittings generate unequal equilibrium abundances, which are imprinted in the relic distribution when inter-sector conversions decouple. This asymmetric distribution retains a memory of the initial axion value and produces an unsuppressed finite-density potential that holds the field at this value until late times. After the WIMP density has sufficiently diluted, the confinement potential drives the axion toward the nearest vacuum. Along this trajectory, the relic-weighted WIMP mass increases monotonically, transferring energy from DE to DM. An observer assuming separately conserved dark components consequently infers an effective DE equation of state that can cross the phantom divide, even though the underlying scalar is canonical and the theory respects the NEC.

The multiplicity of WIMP species also leads to characteristic phenomenological consequences. For universal microscopic scattering cross sections, the total direct-detection rate remains approximately unchanged, whereas indirect-detection signals depend on how broadly the relic abundance is distributed among the sectors and may be enhanced relative to the single-species case.

The late-time DM--DE interaction can additionally suppress structure growth at the percent level. The cosmological solutions presented here are illustrative. A quantitative assessment requires a complete likelihood analysis of both the background evolution and that of cosmological perturbations, jointly fitting CMB temperature, polarization and lensing, integrated Sachs-Wolfe effect, BAO, SNe, and LSS data. 

More broadly, this construction illustrates how an exact microscopic symmetry can protect DE from radiative corrections induced by DM--DE interactions, while an asymmetric DM relic distribution stores the initial axion background value and releases this information billions of years later, generating an apparent phantom crossing of evolving DE.

\section*{Acknowledgment}
We thank Genevi\`eve B\'elanger, Marco Costa and Pasquale Dario Serpico for fruitful discussions. The work of BY is supported by the NSF grant PHY-2309456. SL is supported in part by the Samsung Science Technology Foundation under Project No. SSTF-BA2201-06. This research received no specific funding from CNRS, the ANR, or any other European funding agency.

%%%%%%%%%%%%%%%%%%%%%%%%%%%%%%%%%%%%%%%%%%%%%%%%%%%%%%%%%%%%%%%%%%%
\appendix

%%%%%%%%%%%%%%%%%%

\section{Breaking $\mathbb{Z}_N$ with Asymmetric Reheating}
\label{app:asymmetricReheating}

We consider an alternative realization in which the entire SM is replicated together with the fermion $\chi$, with identical $\chi$--SM interactions in all sectors. These interactions therefore preserve the $\mathbb{Z}_N$ permutation symmetry. Cosmological bounds on additional radiation, conventionally expressed in terms of $N_{\rm eff}$, prevent all SM copies from being populated with comparable temperatures. Reheating must consequently select one sector and thereby break $\mathbb{Z}_N$.

We show that, in a generic realization, the same spurion that produces asymmetric reheating also induces an unsuppressed $\mathbb{Z}_N$-breaking contribution to the PNGB vacuum potential. The requirements of thermal WIMP production and a sufficiently small DE potential are then incompatible unless the reheated SM sector is extraordinarily well sequestered from $\chi$.

As a simple realization, consider a scalar reheaton $\sigma$ coupled preferentially to the Higgs doublet of the $k=0$ sector,
\begin{equation}\label{eq:reheatonL}
    \mathcal{L}_\sigma\supset -m_\sigma^2\frac{\sigma^2}{2}-\kappa_{\rm SM}\sigma H_0^\dagger H_0\,. 
\end{equation}
This super-renormalizable interaction provides the simplest soft source of $\mathbb{Z}_N$ breaking capable of reheating a single SM copy. The limit $\kappa_{\rm SM}\to0$ decouples $\sigma$ from the SM sectors, so a small value of $\kappa_{\rm SM}$ is technically natural~\cite{Hook:2023pba}.

For $m_\sigma\gg m_H$, the decay width of $\sigma$ into a pair of Higgs doublets is 
\begin{equation}
    \Gamma_{\sigma\to {\rm SM}}=\frac{|\kappa_{\rm SM}|^2}{8\pi m_\sigma}\,,
\end{equation}
up to corrections of order $m_H^2/m_\sigma^2$.
Equating this width to the Hubble rate during radiation domination  fixes the reheating temperature, 
\begin{equation}\label{eq:Trh-kappa}
    T_{\rm rh}\simeq 4.6\,{\rm MeV}\sqrt{\frac{m_\sigma}{M_{\rm Pl}}}\left(\frac{|\kappa_{\rm SM}|/m_\sigma}{10^{-20}}\right)\left[\frac{10.75}{g_\star(T_{\rm rh})}\right]^{1/4}\,.
\end{equation}
Successful BBN requires $T_{\rm rh}\gtrsim10\,\MeV$. More importantly for thermal WIMP production, the reheated sector must reach temperatures above the $\chi$ mass, $T_{\rm rh}\gtrsim m_\chi\simeq M$, implying a lower bound on the reheaton coupling,
\begin{equation}
    \frac{|\kappa_{\rm SM}|}{m_\sigma}\gtrsim 3.7\times 10^{-16}\sqrt{\frac{M_{\rm Pl}}{m_\sigma}}\left(\frac{T_{\rm rh}}{M}\right)\left(\frac{M}{100\,{\rm GeV}}\right)\,,
\end{equation}
where we assumed $g_\star(T_{\rm rh})=106.75$.\\

The asymmetric reheating spurion also distinguishes the $\chi$ mass parameter in the populated sector. We parametrize the resulting splitting as $M_0=M(1+\epsilon_b)$ with $|\epsilon_b|\ll1$. At first order in both $\epsilon_b$ and $\epsilon=m_\chi/M$, the CW potential in~\cref{eq:CW} receives the field-dependent correction
\begin{equation}
V_b(\Theta)=-\epsilon_b\frac{\epsilon M^4}{4\pi^2}\left[3\log\frac{M^2}{\mu^2}-1\right]\cos(\Theta-\bar\theta)+\mathcal{O}(\epsilon_b\epsilon^2)\,.
\end{equation}
The sign of the potential is not fixed by the infrared calculation, because the explicit $\mathbb{Z}_N$ breaking reintroduces sensitivity to unknown UV dynamics.  Irrespective of its sign, its natural size should not exceed the observed DE density $\sim\meV^4$. Up to an order-one matching coefficient, this implies
\begin{equation}\label{eq:ebDEconstraint}
|\epsilon_b|\lesssim10^{-53}\left(\frac{0.1}{\epsilon}\right)\left(\frac{100\,\GeV}{M}\right)^4\,.
\end{equation}

The precise relation between $\epsilon_b$ and $\kappa_{\rm SM}$ depends on how the SM communicates with $\chi$. Nevertheless, it is generally proportional to the reheaton VEV induced by the tadpole operator in~\cref{eq:reheatonL} after electroweak symmetry breaking, 
\begin{equation}\label{eq:sigmatadpole}
    \langle\sigma\rangle\simeq -\frac{\kappa_{\rm SM}v^2}{2m_\sigma^2}\,,
\end{equation}
where $v\simeq 246\,\GeV$ is the SM Higgs VEV. As an optimistic naive estimate, one can assume that $\epsilon_b$ ``measures" $\langle\sigma\rangle$ in units of the heaviest scale of the SM, namely $v$, and that the $\chi$ mass is separated from the Higgs sector by $\ell$ loops of SM fields. The induced fractional splitting is then expected to be at least of order $\epsilon_b\sim  L^\ell\langle \sigma\rangle /v$, where $L\equiv g_{\rm SM}^2/(16\pi^2)$ with $g_{\rm SM}\sim 0.1$ denoting a generic coupling of the SM, is a SM loop factor. Using~\cref{eq:Trh-kappa} and imposing $T_{\rm rh}>M$ yields parametrically
\begin{equation}\label{eq:eb-lowerbound}
    |\epsilon_b|\gtrsim 10^{-32-3\ell}\left(\frac{L}{10^{-3}}\right)^\ell\left(\frac{T_{\rm rh}}{M}\right)\left(\frac{M}{100\,\GeV}\right)\left(\frac{M_{\rm Pl}}{m_\sigma}\right)^{3/2}\,,
\end{equation}
where the numerical coefficient is understood at the order-of-magnitude level.

Even in the maximally optimistic limit $m_\sigma\sim M_{\rm Pl}$, reconciling~\cref{eq:eb-lowerbound} with~\cref{eq:ebDEconstraint} requires approximately seven loops for $L\sim 10^{-3}$. For a TeV-scale reheaton, the discrepancy strengthens by about 23 orders of magnitude. Thus asymmetric reheating fails in any generic WIMP realization with ordinary SM-mediated communication between the Higgs and $\chi$. Evading this conclusion would require an additional symmetry or an exceptionally strong sequestering mechanism, beyond the soft $\mathbb{Z}_N$--breaking reheaton coupling itself.

%%%%%%%%%%%%%%%%%%%%%%%%%
\section{Finite-Temperature Potential}\label{app:finiteT}
Here, we compute the finite-temperature axion potential arising from  the fermionic WIMPs when they are in thermal equilibrium with the SM plasma. In the thermal bath of $\chi_k$, the shift-symmetry breaking term in~\cref{eq:Lag} induces a finite-temperature contribution to the axion potential
\begin{align}
V_T(\Theta) = - \frac{2T^4}{\pi^2} \sum_{k=0}^{N-1} J_F\left(u_k^2\right)\,,\quad u_k\equiv \frac{|m_k(\Theta)|}{T}\,. 
\end{align}
The fermionic thermal function~\cite{Dolan:1973qd,Weinberg:1974hy}
\begin{align}
J_F(u^2) \equiv \int_0^\infty {\rm d}y\, y^2 \log\left[1+e^{-\sqrt{y^2 + u^2}}\right]\,,    
\end{align}
admits the high-temperature series expansion~\cite{Quiros:1999jp}
\begin{align}
J_F(u^2) =\,& \frac{7\pi^4}{360} - \frac{\pi^2}{24} u^2 - \frac{u^4}{32} \log\frac{u^2}{a_f} \nonumber\\
&- \frac{\pi^{7/2}}{4}\sum_{\ell = 1}^\infty (-1)^{\ell} \frac{\zeta(2\ell+1)}{(\ell+1)!}\nonumber\\
& \times\left(1-2^{-2\ell -1}\right)\Gamma\left(\ell+\frac{1}{2}\right)\left(\frac{u^2}{\pi^2}\right)^{\ell+2},
\end{align}
where $a_f \equiv \pi^2 \exp\left(3/2-2\gamma_{\rm E}\right)$ with $\log a_f \simeq 2.635$ and $\gamma_{\rm E}\simeq 0.58$ is the Euler-Mascheroni constant, and $\zeta$ is the Riemann $\zeta$-function.

Owing to the $\mathbb Z_N$ symmetry, the ${\cal O}(m_k^2T^2)$ contribution becomes independent of $\phi$ after summing over the $N$ fermion sectors. Moreover, the ${\cal O}(m_k^4\log m_k^2)$ dependence of the thermal correction cancels against the corresponding term in the zero-temperature CW potential. The first field-dependent contribution therefore arises at ${\cal O}(\epsilon^N)$. In the high-temperature regime, $M\ll T$, its leading form at large $N$ is
\begin{align}
V_T(\Theta) \simeq  (-1)^N\frac{\pi^{3/2}T^4}{\sqrt{N}} \epsilon^N\left(\frac{M}{\pi T}\right)^{2N}   \cos\left[N(\Theta-\bar\theta)\right]\,. 
\end{align}
This contribution is extremely suppressed, not only by $\epsilon^N$ but also by the high-temperature factor $\left[M/(\pi T)\right]^{2N}$, and is therefore much smaller than the confining potential in~\cref{eq:Vconf}. Consequently, for $T\gg M $, the axion remains frozen at the value inherited from inflation.\\

In the low-temperature regime, $T\ll M$, the fermionic thermal function admits a Boltzmann expansion in $e^{-n\sqrt{u}}$, with each term accompanied by an asymptotic series in $1/\sqrt{u}$. The leading $n=1$ contribution scales as $u^{3/4}e^{-\sqrt{u}}$, while higher terms are exponentially suppressed by $e^{-n\sqrt{u}}$ and encode Fermi-Dirac corrections. Thus, the finite-temperature potential becomes exponentially small once the fermions are nonrelativistic. 
Also in this regime, the axion potential is dominated by the vacuum contribution, and the field stays frozen at its inflationary value.

%%%%%%%%%%%%%%%%%%%%%%%%%
\section{Finite-Density Potential}\label{app:fdpotential}
We derive here the finite-density potential generated by the nonuniform relic distribution among the $N$ fermion sectors. 
The relative mass splittings at leading order in $\epsilon$ is
\begin{equation}
    \Delta_k \simeq \epsilon\left[\cos\left(\Theta_i -\bar\theta+\frac{2\pi k}{N}\right)+\cos\delta\right]+\mathcal{O}(\epsilon^2)\,.
\end{equation}
Working at leading order in $\epsilon$, while keeping $\epsilon x_d$ unexpanded in the Boltzmann factor, and neglecting corrections of order $\epsilon^2x_d$, the abundance fractions are 
\begin{equation}
    \eta_k\simeq \frac{e^{-q \cos(\Theta_i-\bar\theta+ 2\pi k/N)}}{N Z_N(q,\Theta_i)}+\mathcal{O}(\epsilon^2,\epsilon^2q)\,,
\end{equation}
where
\begin{equation}
    Z_N(q,\Theta_i)\equiv \frac{1}{N}\sum_{k=0}^{N-1}e^{-q\cos(\Theta_i-\bar\theta+2\pi k/N)}\,.
\end{equation}

The finite-density potential after conversion decoupling may then be written as
\begin{align}
    V_{\rm fd}(\Theta)\simeq&\, n_\chi M \left[1+\epsilon\sum_{k=0}^{N-1}\cos\left(\Theta-\bar\theta+\frac{2\pi k}{N}\right)\eta_k+\mathcal{O}(\epsilon^2)\right] \nonumber\\
    =&\, n_\chi M\Bigg[1-\epsilon \bigg[\partial_q \log Z_N(q,\Theta_i) \cos(\Theta-\Theta_i)\nonumber\\
    &\quad\quad\, +\frac{1}{q}\,\partial_{\Theta_i} \log Z_N(q,\Theta_i) \sin(\Theta-\Theta_i)\bigg]\Bigg]\,.
\end{align}
The function $Z_N(q,\Theta_i)$ admits a decomposition on the modified Bessel functions of the first kind,
\begin{equation}
Z_N(q,\Theta_i)=I_0(q)+2\sum_{\ell\geqslant1} (-1)^{N\ell} I_{N\ell}(q)\cos[N\ell(\Theta_i -\bar\theta)]\,,   
\end{equation}
as follows from the root-of-unity sum. In the large $N$ limit  at fixed $q$, only the $I_0$ term remains, since $I_{N\ell}(q)\simeq (q/2)^{N\ell}/(N\ell)!\to 0$. As a result, the finite-density potential becomes
\begin{equation}
    V_{\rm fd}(\Theta)\simeq n_\chi M\left[1-\epsilon\frac{I_1(q)}{I_0(q)}\cos\left(\Theta-\Theta_i\right)\right]\,.
\end{equation}

%%%%%%%%%%%%%%%%%%%%%%%%%
\section{Linear Cosmology}\label{app:linpert}
Working in the conformal Newtonian gauge and assuming negligible anisotropic stress, the evolution of the DM density contrast $\delta_\chi(\boldsymbol{r},t)\equiv\rho_\chi(\boldsymbol{r},t)/\bar \rho_\chi(t)-1$, velocity divergence $\theta_\chi$, and axion perturbation $\delta\phi(\boldsymbol{x},t)\equiv\phi(\boldsymbol{x},t)-\bar\phi(t)$ in Fourier space with comoving wavenumber $\boldsymbol{k}$ is governed at linear order by
\begin{align}
    \delta_\chi'+\theta_\chi-3\Psi'-\delta\phi'\beta(\bar\phi)-\bar\phi'\delta\phi\frac{{\rm d}\beta}{{\rm d}\bar\phi}&=0\,,\\
    \theta_\chi'+\left[\mathcal{H}+\beta(\bar\phi)\bar\phi'\right]\theta_\chi-k^2\left[\Psi+\beta(\bar\phi)\delta\phi\right]&=0\,,
\end{align}
where barred quantities denote homogeneous background values, $\Psi$ is the Newtonian gravitational potential, and $\delta\phi$ satisfies the perturbed Klein-Gordon equation
\begin{widetext}
\begin{equation}
\delta\phi''+2\mathcal{H}\delta\phi'-4\bar\phi'\Psi'+\delta\phi\left[k^2+a^2\left(\frac{{\rm d}^2V}{{\rm d}\bar\phi^2}+\bar\rho_\chi\frac{{\rm d}\beta}{{\rm d}\bar\phi}\right)\right]
+2\Psi a^2\left[\frac{{\rm d}V}{{\rm d}\bar\phi}+\bar\rho_\chi\beta(\bar\phi)\right]+a^2\bar\rho_\chi \delta_\chi \beta(\bar\phi)=0\,.
\end{equation}
\end{widetext}
The term proportional to \(\beta\delta\phi\) is the scalar fifth force,
while the term \(\beta\bar\phi'\theta_\chi\) is an additional drag induced by the time dependence of the WIMP mass. In the
phantom-crossing regime of interest, the DM mass typically increases at
late times, so \(\beta\bar\phi'>0\), and this term acts as an extra
friction on the DM velocity.

These equations are supplemented by the usual perturbation equations for
photons and baryons, together with the Einstein equations for the
metric perturbation \(\Psi\). The expressions above assume that the
pressure and velocity dispersion of the WIMPs
can be neglected on the large scales relevant for the late-time growth
analysis.

To solve the linear perturbation equations, we impose adiabatic initial conditions at $z_i=10^7$ coming from inflation. At this redshift, all Fourier modes relevant for the computation of the matter power spectrum satisfy $k\ll\mathcal H$ and are therefore outside the cosmological horizon. The primordial scalar perturbations from inflation are assumed to be nearly scale invariant, with amplitude $A_s$ and spectral index $n_s$ fixed to their Planck 2018 best-fit values~\cite{Planck:2018vyg}. Since $\bar\phi'(z_i)=0$, adiabaticity further requires that the axion perturbations vanish initially, yielding $\delta\phi(z_i)=\delta\phi'(z_i)=0$.
The remaining perturbation variables are therefore initialized according to the standard adiabatic conditions of the $\Lambda$CDM cosmology. Since the axion field is frozen at early times, the evolution of cosmological perturbations remains essentially indistinguishable from that of $\Lambda$CDM until the onset of the late-time axion dynamics.

%%%%%%%%%%%%%%%%%%%%%%%%%%%%%%%
\bibliographystyle{JHEP}
\bibliography{refs.bib}

\end{document}